\documentclass[12pt,a4paper]{article}
\begin{document}
\textwidth=135mm
 \textheight=200mm
\begin{center}
{\bfseries The inaction approach to gauge theories}
\vskip 5mm
G.B. Pivovarov
\vskip 5mm
{\small {\it Institute for
Nuclear Research, 117312 Moscow, Russia}} 
\end{center}
\vskip 5mm
\centerline{\bf Abstract}
The inaction approach introduced previously for $\phi^4$ \cite{gbp} is generalized to gauge theories. It combines the advantages of the effective field theory and causal approaches to quantum fields. Also, it suggests ways to generalizing gauge theories.
\vskip 10mm
\section{\label{sec:intro}Introduction}

Elementary particles are described by the standard model. The standard model is a gauge theory. While being a phenomenological success, it is not generally considered as a satisfactory theory. This is the case because the standard model requires several tens of input parameters, and, on top of this, the parameters should be fine tuned (the naturalness problem). 

Attempts to improve the standard model are unsuccessful for the last thirty years. A view has been formed that the standard model is only a low energy approximation to a more satisfactory (yet unknown) theory which should include quantum gravity.

On a more technical level, this view is partially realized within the effective field theory approach to quantum fields \cite{effective}. Withing this approach, UV divergences and UV regularization obtain a physical interpretation. From now on I call the combination of the above view with the effective field theory approach  the \textit{standard approach}. 

The observations at the LHC are currently at odds with the standard approach: no superpartners, no extra dimensions, no exotics. We are left with the unsatisfactory but successful standard model.

Accepting the situation, I suggest to reconsider our approach to gauge theories. The assumption behind this suggestion is that the attempts at grand unification may have failed because of the language that has been used. The key words of this language are ``action" and ``lagrangian". For gauge theories, the action should be gauge invariant. Thus, within the standard approach, the game is to suggest a gauge invariant local action and see the consequences.

This standard approach has an obvious flaw: its key object---the action---does not exist without a regularization. This flaw is particularly striking for the chiral models, where the bare action is not even gauge invariant because of the absence of a regularization preserving the $\gamma_5$ symmetries \cite{chiral}.

On the other hand, the advantages of the standard approach are that we are used to it, and that it has deep historical roots. At this point, I want to stress that the approach sketched here is an extension of the standard approach: all that can be said in the standard way can also be expressed in the language developed here. Hopefully, the reverse is not true: there is a gain in talking the new language. And the price of switching to this new language is not prohibitively high.

I start characterization of the new approach with a negation. It does not use the action of the system as a building block. Because of this, from now on, I call the new approach the \textit{inaction approach}.

An approach already exists that avoids functional methods in general and the action functional in particular. This is the causal approach of Epstein-Glaser-Scharf \cite{scharf}. In this approach, one attempts to avoid using unobservable objects like Green functions. The key object here is the $S$-matrix. The $S$-matrix of the causal approach is an operator-valued functional depending on the couplings which in turn are functions of the spacetime location. The leading nontrivial terms in the expansion of the $S$-matrix in powers of the couplings are local operators that parameterize the theory.

In my view, the causal approach is idiosyncratic. It rejects the use of Green functions. Outside the causal approach, Green functions are used extensively. The experience with perturbative QCD teaches us that extraction of observables from Green functions is an involved process. It uses knowledge gained in the computation of Green functions. So, one should not rush with defining observables of the theory. To an extent, a theory  itself defines its observables. Defining observables is a task which may be deferred.

The inaction approach advocated here studies Green functions and employs functional integration. In this respect it is closer to the standard approach than the causal approach. So, the inaction approach lies between the causal approach and the standard approach. It shares with the causal approach the advantage of being regularization free: no UV divergences appear if one is careful enough. At the same time, like the standard approach, the inaction approach is not shy of using unobservable objects.

In Section \ref{sec:basics}, I introduce key objects of the inaction approach to gauge theories. The presentation here is informal. I intermix motivations and definitions. New terms appear first in \textit{italic type}. 

In Section \ref{sec:inact}, I give details on the inaction equation---the key equation of the inaction approach. In Section \ref{sec:super-t}, I describe the super translation invariance, which is the way the gauge invariance appears within the inaction approach. 
In Section \ref{sec:out}, I give my conclusions and outlook. 

\section{\label{sec:basics}The inaction basics}

Withing the inaction approach, a theory with all input parameters fixed is represented by a generating functional of connected Green functions $W(J)$ ($J$ here is the set of sources for the fields of the model). It is assumed that $W(J)$ is translation invariant. Apart of that, no restrictions on $W(J)$ are assumed at this stage. So, a theory $W(J)$ is a vector in the infinite dimensional linear space of translation invariant  connected functionals of a chosen set of sources $J$. This space is the \textit{Green space} $\mathcal{G}$. The connectedness and translation invariance means that any homogeneous term in the expansion of $W\in\mathcal{G}$ in powers of the sources is a convolution of the sources with a Fourier transform of a product of the delta function expressing the momentum conservation and a sufficiently smooth Green  function of the momenta.

Without loss of generality, I take that $W(J)=O(J^2)$, which means that $J$ are the sources to the deviations of the fields from the vacuum values.

The first task is to restrict a theory $W\in\mathcal{G}$ in such a way that it would correspond to a local action of a renormalizable theory. One needs to do this without using a particular local action, because, as known, the action should be infinite to make $W$ nontrivial and finite.

This task is accomplished with the \textit{inaction equation} for a local renormalizable theory $W$. The inaction equation reads
\begin{equation}
\label{inaction}
W=L_q^{-1}\circ P_\mu\circ L_q[W]
\end{equation}
 Here $L_q$ is a \textit{quantum Legendre transform}, $P_\mu$ is a projector onto a finite dimensional linear space of local connected functionals, and $L_q^{-1}$ is the inverse quantum Legendre transform.

The quantum Legendre transform is a nonlinear mapping that eats $W(J)$ and spits the action of the system. In the tree approximation, quantum Legendre is just the familiar Legendre transform. The quantum Legendre can be expressed with functional integration (see Section \ref{sec:inact}).

The projector $P_\mu$ is a linear function on the space of translation invariant connected functionals. Its range is a finite dimensional linear space. If a theory $W$ corresponds to a local renormalizable action, $P_\mu\circ L_q[W]=L_q[W]$, which implies the inaction equation (\ref{inaction}). The space of such $P_\mu$ is infinite dimensional. I will point out particular $P_\mu$ parameterized with a finite set of \textit{normalization parameters} $\mu$ (see Section \ref{sec:inact}).

For brevity, I define the \textit{inaction mapping} $I_\mu\equiv L_1^{-1}\circ P_\mu\circ L_q$. With this notation, the inaction equation is $W=I_\mu[W]$.

This equation is a fixed point equation satisfied by any local renormalizable theory $W$. Its solutions form a finite dimensional surface embedded in the Green space. This surface of fixed points constitutes a \textit{theory surface} $\mathcal{T}\subset\mathcal{G}$.  Particular points on $\mathcal{T}$ correspond to particular theories with all input parameters of the theory fixed, and the input parameters of the theory are the coordinates on the surface of solutions to the inaction equation. 

The inaction approach tries to study the properties of the theory surface $\mathcal{T}$, and to describe physics with these properties.

In \textit{perturbation theory}, one studies local properties of $\mathcal{T}$ near a particular solution to the inaction equation (\ref{inaction}) corresponding to a \textit{free theory}. For a free theory, $W=W_F\in \mathcal{T}$, where $W_F$ is quadratic in the sources. Linearizing the inaction equation near $W_F$ one obtains
\begin{equation} 
\label{inaction-l}
W = W_F+\tilde{P}_\mu (W-W_F) + O\Big((W-W_F)^2\Big),
\end{equation}
where $\tilde{P}_\mu$ is the linear part of the inaction mapping $I_\mu$ near the quadratic fixed point $W_F$.

If I apply the inaction mapping to both sides the above linearized equation and again linearize it in the right hand side, I obtain that $\tilde{P}_\mu$ is a projector. If it is a not a unit operator, it nullifies a nontrivial subspace. Geometrically, the range subspace of $\tilde{P}_\mu$ is the tangent space to the theory surface $\mathcal{T}$ at the point $W_F$.

If $L_q$ is non degenerate at $W_F$, $\tilde{P}_\mu$ is similar to $P_\mu$ from the definition of the inaction mapping, with the similarity transformation defined by the differential of $L_q$ at $W_F$. 

Introducing a notation, $\tilde{P}_\mu$ is a projector onto a finite dimensional linear \textit{root space} $\mathcal{R}$ which is tangent to $\mathcal{T}$ at the point $W_F$. The root space is a finite dimensional linear subspace of the Green space, $\mathcal{R}\subset\mathcal{G}$. The projection of $W\in \mathcal{T}$ onto $\mathcal{R}$, $\tilde{P}_\mu (W-W_F)\equiv R_\mu\in \mathcal{R}$, is the \textit{root} of the theory $W\in \mathcal{T}$.

A theory $W\in\mathcal{T}$ can be expanded in powers of its root $R_\mu$, with the free theory $W_F$ and the root $R_\mu$ being  the first two terms of the expansion:
\begin{equation}
\label{expansion}
W=W_F+R_\mu +\sum_{k=2}^\infty W_n(R_\mu),
\end{equation}
where $W_n$ are homogeneous in $R_\mu$: $W_n(\lambda R_\mu)=\lambda^n W_n(R_\mu)$. I will explain how the inaction equation (\ref{inaction}) fixes uniquely the homogeneous functions $W_n$ (see Section \ref{sec:inact}).

The expansion (\ref{expansion}) constitutes  the perturbation theory of the inaction approach, and the root $R_\mu$ replaces the action of interaction of the standard approach. Importantly, $R_\mu$ is finite along with the $W_n$, if the projector $P_\mu$ in (\ref{inaction}) corresponds to a renormalizable theory. UV divergences do not appear. If one accepts the inaction approach, UV divergences and UV regularizations become artifacts. 

Notice that the root $R_\mu$ depends on the normalization point $\mu$. This dependence should be such that the theory $W$ would be independent of $\mu$. Requiring this one obtains \textit{renormalization group equations} for the root $R_\mu$.

This program has been realized in \cite{gbp} for $\phi^4$. Now I will describe the extension needed to include gauge theories.

First of all, why one should look for an extension? An extension is needed because the inaction equation is not restrictive enough. There are theories among its solutions that do not admit physical interpretation. A notable example is provided by the Curci-Ferrari model which is local, renormalizable, but nonunitary \cite{curci}.

To have a chance for physical interpretation, one should further restrict the theory requiring from a physically viable theory $W$ that it would satisfy some form of the Slavnov-Taylor identities.

To impose the Slavnov-Taylor, it is convenient to extend the set of sources $J$ with the sources for the generators of the BRST-antiBRST symmetry \cite{bal}. In this way, each source component becomes quadrupled, because for each field component there are generators of its BRST, antiBRST, and mixed BRST-antiBRST transformations.

These new sources can be described very economically if one adds two Grassmann coordinates to the spacetime, taking that each source depends not only on the conventional spacetime coordinates, but also on the new Grassmann coordinates. Expansion of each source component in powers of the introduced Grassmann coordinates generates the needed extra components of the sources for the generators of the BRST-antiBRST transformations.

It turns out that the identities for $W$ corresponding to BRST-antiBRST symmetry are equivalent to requiring that $W$ be translation invariant not only with respect to shifts in the conventional spacetime, but also with respect to shifts along the two extra Grassmann coordinates.

This is detailed in Section \ref{sec:super-t}. In particular, I discuss there the modification of the quantum Legendre transform from the inaction equation (\ref{inaction}) which is required because not all the components of the sources are involved in the transformation after the extension of the set of sources.  

What matters at this stage of the presentation is that gauge invariance constitutes extra conditions on the theory $W$, and these conditions are linear. I will call the theories satisfying these conditions \textit{super translation invariant}, and will denote the space of super translation invariant theories by $\mathcal{I}$. For example, the expression $W\in \mathcal{I}$ means that $W$ is super translation invariant. I stress that $\mathcal{I}\subset\mathcal{G}$ is a linear subspace of the Green space: a linear combination of super translation invariant theories is a super translation invariant theory.

Because there exist local renormalizable gauge theories, there exist super translation invariant theories satisfying the inaction equation. In other words, the intersection of the surface of local renormalizable theories $\mathcal{T}$ and the space of super translation invariant theories $\mathcal{I}$ is nonepmty: $\mathcal{T}\cap\mathcal{I}\equiv\mathcal{P}\neq\emptyset$. Here $\mathcal{P}$ denotes a finite dimensional surface of \textit{physically viable theories}, which are super translation invariant local renormalizable theories.

Now I want to describe a perturbation theory for $W\in\mathcal{P}$. Let $W_F\in\mathcal{P}$ be a physically viable free theory, which means that it is quadratic in the sources, satisfies the inaction equation, and super translation invariant. I want to parameterize $\mathcal{P}$ near $W_F$.

To achieve this, consider the intersection of the root space $\mathcal{R}$ corresponding to the super translation invariant free theory $W_F$ with the space of super translation invariant theories $\mathcal{I}$. This is the \textit{seed space} $\mathcal{S}=\mathcal{R}\cap\mathcal{I}$. As an intersection of the two linear spaces one of which is finite dimensional, $\mathcal{S}$ is a finite dimensional linear space.

Evidently, the seed space $\mathcal{S}$ is the tangent space to the surface of the physically viable theories $\mathcal{P}$ at the point $W_F$. This is the case because $\mathcal{R}$ is tangent to $\mathcal{T}$ at $W_F$, and $\mathcal{P}=\mathcal{T}\cap\mathcal{I}$.

The seed space $\mathcal{S}$ contains all the solutions to the linearized inaction equation (\ref{inaction-l}) that satisfy the condition of super translation invariance. While local renormalizable theories are parameterized by the coordinates in the root space $\mathcal{R}$ (see Eq. (\ref{expansion})), the physically viable theories are parameterized by the coordinates in the seed space $\mathcal{S}\subset\mathcal{R}$. 

If the projector $\tilde{P}_\mu$ would act along the space $\mathcal{I}$, it would project the surface of physically viable theories $\mathcal{P}$ onto the seed space $\mathcal{S}$. In this case, the expansion (\ref{expansion}) would be the perturbation theory for a physically viable theory $W\in\mathcal{P}$ if the root $R_\mu$ in the right hand side of (\ref{expansion}) would be taken from the seed space, $R_\mu\in\mathcal{S}$.

But this is generally not the case. $\tilde{P}_\mu$ projects $\mathcal{P}-W_F$ onto a \textit{surface of physical roots} $\mathcal{P}_\mu\equiv P_\mu(\mathcal{P}-W_F)$ embedded into the finite dimensional root space. The dimension of this surface equals the dimension of the seed space $\mathcal{S}$. 

I conclude that the perturbative parameterization of the $\mathcal{P}$ near $W_F$ may be achieved in two stages. On the first stage, one finds the perturbative parameterization of $\mathcal{P}_\mu$ in terms of the coordinates in the seed space. On the second stage, the obtained parameterization of the roots is substituted in (\ref{expansion}) and we obtain the desired parameterization of a physically viable theory in terms of the coordinates in the seed space.

Intuitively, renormalized couping of a three gluon vertex is a coordinate in the seed space. Coordinates in the seed space are independent parameters of the theory. The four gluon coupling is a coordinate on $\mathcal{P}_\mu$. It is not independent and can be expressed in terms of the three gluon coupling.

To describe the location of $\mathcal{P}_\mu$ in $\mathcal{R}$, I need equations for $R_\mu\in\mathcal{P}_\mu$. These equations should guarantee that any $R_\mu$ satisfying them would yield a super translation invariant $W$ in the left hand side of (\ref{expansion}).

Here are these equations:
\begin{equation}
\label{super-t-for-r}
s_iR_\mu= -\sum_{n=2}^\infty s_i W_n(R_\mu).
\end{equation}
The subscript $i=1,2$ numbers the translations along the two Grassmann variables discussed above. The action of the super translation generators $s_i$ is described explicitly in Section \ref{sec:super-t}.

These equations are obtained from Eq. (\ref{expansion}) by acting on both sides of the equation with the super translation generators $s_i$, and taking into account that both $W$ and $W_F$ are super translation invariant, $s_i W= s_i W_F = 0$.

Eq. (\ref{super-t-for-r}) demonstrates that the roots corresponding to super translation invariant theories are not super translation invariant. But the breaking of the super translation invariance in $R_\mu$ is uniquely defined by the invariance of $W$ and $W_F$.

The system of equations (\ref{super-t-for-r}) is overdetermined: this is an infinite number of equations for finite number of coordinates of $R_\mu$. In the linear approximation, the right hand side of these equations can be dropped, and I obtain the equations defining the seed space $\mathcal{S}$. At the moment, I do not understand the mechanism which guarantees the existence of nontrivial solutions to (\ref{super-t-for-r}). But I know that solutions exist, because gauge theories do exist. These solutions form the surface $\mathcal{P}_\mu$.  

The parameterization of $\mathcal{P}_\mu$ with locations in the seed space is further discussed in Section \ref{sec:super-t}. In particular, I argue there that a unique \textit{seed} $S_\mu\in\mathcal{S}$ corresponds to any super translation invariant theory $W\in\mathcal{P}$, and that there exists an expansion of $W$ in powers of $S_\mu$ in which $W_F$ and $S_\mu$ constitute the zero and the first term, in analogy with the expansion (\ref{expansion}). 

I was to introduce quite a number of notions in this section. See the Table for a short inaction vocabulary.

\begin{table}
\centering
\caption{Inaction vocabuary}
\begin{tabular}{c | p{8cm}}
\hline
notion &meaning\\
\hline\hline
Green space $\mathcal{G}$ &An infinite dimensional linear space of generating functionals for translation invariant connected Green functions.\\
\hline
Theory $W$& A vector in $\mathcal{G}$, $W\in\mathcal{G}$.\\
\hline
Inaction mapping $I_\mu\equiv L_q^{-1}\circ P_\mu\circ L_q$& The mapping in the right hand side of the inaction equation $W=I_\mu[W]$.\\
\hline 
Theory surface $\mathcal{T}\subset\mathcal{G}$ &A finite dimensional surface of solutions to the inaction equation. Each point of this surface corresponds to a local renormalizable theory.\\
\hline
Free theory $W_F\in\mathcal{T}$&A local renormalizable theory quadratic in the sources.\\
\hline
Root space $\mathcal{R}\subset\mathcal{G}$ & A finite dimensional subspace of $\mathcal{G}$. Tangent to $\mathcal{T}$ at $W_F$. Parametrizes $\mathcal{T}$ near $W_F$\\ 
\hline
Theory root $R_\mu\in\mathcal{R}$ & A point in $\mathcal{R}$ corresponding to a particular local renormalizable theory $W\in\mathcal{T}$ that is close to a free theory $W_F\in\mathcal{T}$: $W = W_F+R_\mu+O(R_\mu^2)$.\\
\hline 
Subspace of invariants $\mathcal{I}\subset\mathcal{G}$ & A linear infinite dimensional subspace of the Green space $\mathcal{G}$. Contains super translation invariant theories.\\
\hline 
Physical theory surface $\mathcal{P}=\mathcal{T}\cap\mathcal{I}$ &The surface of physically viable theories, which are local, renormalizable, and super translation invariant.\\
\hline
Seed space $\mathcal{S}=\mathcal{R}\cap\mathcal{I}$ &Finite dimensional linear  space. Tangent to $\mathcal{P}$ at a free super translation invariant theory $W_F$. Parameterizes $\mathcal{P}$.\\
\hline
Surface of physical roots $\mathcal{P}_\mu$ & Image of $\mathcal{P}$ in $\mathcal{R}$: $\mathcal{P}_\mu=P_\mu(\mathcal{P}-W_F)$.\\
\hline
Theory seed $S_\mu\in\mathcal{S}$& A point in $\mathcal{S}$ corresponding to a particular super translation invariant local renormalizable theory $W\in\mathcal{P}$ that is close to a free theory $W_F\in\mathcal{P}$: $W = W_F+S_\mu+ O(S_\mu^2)$.\\ 
\hline
\end{tabular}

\end{table}

\section{\label{sec:inact}The inaction equation}

As promised, I detail in this section the quantum Legendre transform $L_q$, the projector onto the space of local renormalizable actions $P_\mu$ from (\ref{inaction}), and derivation of (\ref{expansion}). 

I remind that $L_q$ acts on $W$ and gives the corresponding action of the system $I$. We know the action of $L_q^{-1}$ on $I$:
\begin{equation}
\label{action}
e^{L_q^{-1}[I](J)}=\int\mathcal{D}\phi\,e^{I(\phi)+J\phi}.
\end{equation}
This is the standard functional integral representation of $W\equiv L_q^{-1}[I]$. As we know, $I$ does not exist without a regularization if $W$ is finite. 

Remarkably, a similar formula exists for $L_q$:
\begin{equation}
\label{DE}
e^{L_q[W](\phi)}=\int\mathcal{D}J\,e^{W(J)-J\phi}.
\end{equation} 
This fact is known as the Dominicis-Englert duality. For a formal proof and some references, see \cite{gbp}.

I conclude that
\begin{equation}
\label{explicit}
L_q[W](\phi)=\hbar\log{\big[\int\mathcal{D}J\,e^{(W(J)-J\phi)/\hbar}\big]},
\end{equation}
where I restored temporarily the Planck constant.

With the standard assumptions of the quasi-classical methods, one derives that the leading term of the expansion of the right hand side of (\ref{explicit}) in powers of the Planck constant is just the Legendre transform:
\begin{equation}
\label{legendre}
L_q[W](\phi)=\max_J{[W(J)-J\phi]}+O(\hbar).
\end{equation}
As promised, the quantum Legendre transform is a quantum generalization of the Legendre transform.

I can also determine the expansion of $L_q[W]$ in powers of $W-W_F$, where $W_F$ is quadratic in the sources:
\begin{equation}
\label{legendre-e}
L_q[W](\phi)=I_F(\phi)+\Big(W-W_F\Big)(-\frac{\delta I_F}{\delta\phi}-\frac{\delta}{\delta\phi})+ O\Big((W-W_F)^2\Big),
\end{equation}
where $I_F$ is the free action $L_q[W_F]=I_F$; the sign by the variational derivatives is given for a commuting source component. For a Grassmann source component the sign should be inverted. 

For the free action above I have $I_F(\phi)=W_F(J[\phi])-J[\phi]\phi$, and $J[\phi]$ is defined by the equation
\begin{equation}
\label{source}
\Big[\frac{\delta W_F}{\delta J}\Big]_{J[\phi]}=\phi.
\end{equation}

I point out that $L_q$ is ill defined: UV divergences appear in $L_q[W]$ if $W$ is finite. But I never use $L_q$ alone. The only combination I use is $I_\mu\equiv L_q^{-1}\circ P_\mu \circ L_q$, which is well defined for $P_\mu$ defined below.

Next I define $P_\mu$ involved in the definition of  the inaction mapping $I_\mu$.
$P_\mu$ should be a projector, $P_\mu^2=P_\mu$, and it should project onto the finite dimensional linear space to which any local renormalizable action belongs, $P_\mu I=I$.

$I$ is a sum of terms homogeneous in the fields, starting from the quadratic term. The first property of $P_\mu$ is that it acts independently on each homogeneous contribution to $I$.

Let us start from defining the action of $P_\mu$ on the term quadratic in the fields. The general form of this term is
\begin{equation}
\label{quadric}
I_2(\phi)=\int d^4k\,I_{\alpha\beta}(k)\phi^\alpha(-k)\phi^\beta(k).
\end{equation}
Here $\alpha$ and $\beta$ index the field components. 

A mass dimension $d_\alpha=1,3/2,2$ is assigned to each field component. For a local theory, $I_{\alpha\beta}(k)$ is a polynomial in the momentum $k$. Its degree does not exceed $4-d_\alpha-d_\beta$.  

$P_\mu$ will act on $I_2$ by replacing the involved tensor $I_{\alpha\beta}(k)$ with a transformed tensor. A possible choice is to replace $I_{\alpha\beta}$ with a piece of its Taylor expansion around an arbitrary four momentum $q_0$, starting with the constant term, through the terms which degrees do not exceed $4-d_\alpha-d_\beta$. Evidently, with this choice, $P_\mu I_2=I_2$ for a local theory. The reference momentum $q_0$ is among the set of the normalization parameters: $q_0\in\mu$. 

Let us do analogously for the term cubic in the fields:
\begin{equation}
\label{cubic}
I_3(\phi)=\int d^4k_1d^4k_2\,I_{\alpha\beta\gamma}(k_1,k_2)\phi^\alpha(-k_1-k_2)\phi^\beta(k_1)\phi^\gamma(k_2).
\end{equation}
In a local renormalizable theory, the terms appear only if $4-d_\alpha-d_\beta-d_\gamma\geq 0$. The tensor $I_{\alpha\beta\gamma}(k_1,k_2)$ should be in this case a polynomial which degree does not exceed $4-d_\alpha-d_\beta-d_\gamma$.

$P_\mu$ will act on $I_3$ by replacing the involved tensor $I_{\alpha\beta\gamma}(k_1,k_2)$ with a transformed tensor. A possible choice is to replace $I_{\alpha\beta\gamma}$ with a piece of its Taylor expansion around an arbitrary four momenta $q_1,q_2$, starting with the constant term, through the terms which degrees do not exceed $4-d_\alpha-d_\beta-d_\gamma$. Evidently, with this choice, $P_\mu I_3=I_3$ for a local renormalizable theory. The reference momenta $q_1,q_2$ are among the set of the normalization parameters: $q_{1,2}\in\mu$. 

Similarly I do for the $I_4$. The difference here is that the only terms of this sort allowed in a local renormalizable theory are constructed from the fields with $d_\alpha=1$ and are independent of the corresponding triple of the four momenta. Correspondingly, $P_\mu$ will replace $I_{\alpha\beta\gamma\delta}(k_1,k_2,k_3)$ with its value at the reference momenta $q_3,q_4,q_5$.

A local renormalizable theory does not contain terms of more than fourth power in the fields. Correspondingly, $P_\mu I_n\equiv 0$ for $n>4$.

I conclude that the set of normalization points consists of the six reference four momenta: $\mu=\{q_0,\dots,q_5\}$. $P_\mu$ replaces the coefficient tensors $I_{\alpha\dots\delta}(k_1,\dots, k_n)$ with the pieces of their Taylor expansions in the momenta around the reference momenta each time it is possible without running into coefficients of the polynomials with negative mass dimensions.

Next I derive the perturbative expansion (\ref{expansion}). To this end, I recall that $\tilde{P}_\mu (W-W_F)=R_\mu$, and rewrite the inaction equation as follows:
\begin{equation}
\label{rewrite}
W=W_F+R_\mu+(1-\tilde{P}_\mu)\Big(I_\mu[W]-W_F-R_\mu\Big).
\end{equation}
Introducing notation $X\equiv (1-\tilde{P}_\mu)(W-W_F)$, I rewrite it one more time:
\begin{equation}
\label{transversal}
X=(1-\tilde{P}_\mu)\Big(I_\mu[W_F+R_\mu+X]-W_F-R_\mu\Big).
\end{equation}

Intuitively, the component of $(W-W_F)$ along the root space $\mathcal{R}$ is $R_\mu$, and $X$---the component of $(W-W_F)$ along the null space of $\tilde{P}_\mu$---is a function of $R_\mu$. Its expansion in $R_\mu$ around the zero point does not have a linear term, because of the structure of the right hand side of (\ref{transversal}): $I_\mu[W_F+R_\mu+X]-W_F-R_\mu=O((R_\mu+X)^2)$.

I conclude that
\begin{equation}
\label{first}
X=(1-\tilde{P}_\mu)\Big(I_\mu[W_F+R_\mu]-W_F-R_\mu\Big)+O(R_\mu^3) .
\end{equation}

From this I deduce that 
\begin{equation}
\label{w2}
W_2(R_\mu)=(1-\tilde{P}_\mu)\Big(I_\mu[W_F+R_\mu]\Big)_2,
\end{equation}
where $W_2$ is the term in the right hand side of (\ref{expansion}). The subscript $2$ in the right hand side means that only the terms quadratic in $R_\mu$ are retained within the brackets.

Generalizing I have
\begin{equation}
\label{w2}
W_n(R_\mu)=(1-\tilde{P}_\mu)
\Big(I_\mu[W_F+R_\mu+W_2(R_\mu)+\dots+W_{n-1}(R_\mu)]\Big)_n.
\end{equation}

The right hand side of (\ref{expansion}) is constructed.

\section{\label{sec:super-t}The super translation invariance}

It is demonstrated in \cite{bal} that gauge theories are uniquely characterized as theories possessing BRST and antiBRST symmetries. The actions of gauge theories are invariant with respect to field transformations of the form
\begin{equation}
\label{transf}
\phi^\alpha\rightarrow\phi^\alpha +\epsilon^1(s_1\phi)^\alpha+\epsilon^2(s_2\phi)^\alpha, 
\end{equation}
where $(s_i\phi)^\alpha$ are certain local polynomials in the fields. For $i=1(2)$, it is the generator of the BRST (antiBRST) transformation, correspondingly. Here the transformation parameters $\epsilon^i$ are Grassmann numbers with mass dimension $-1$. See \cite{bal} for explicit definitions of $s_i\phi$.

I will not need an explicit form of the local polynomials $s_i\phi$ defining the transformation (\ref{transf}). But we have to consider how they are transformed under the transformation (\ref{transf}):
\begin{equation}
\label{iter}
(s_i\phi)\rightarrow (s_i\phi)+\epsilon^{j}\epsilon_{ji}s\phi,
\end{equation}
where $j$ is summed over, $\epsilon_{ji}$ is antisymmetric in the indexes with $\epsilon_ {12}=1$, and $s\phi$ is a new local polynomial of the fields. 

It is important that the new local polynomial of the fields $s\phi$ appearing in (\ref{iter}) is invariant with respect to the transformations (\ref{transf}). 

I conclude that the local polynomials of the fields $s_i\phi, s\phi$ must be important objects, and, therefore, we want to know the Green functions with insertions of any number of the corresponding operators. So, we replace the term $J\phi$ in the exponent under the functional integral of (\ref{action}) in the following way:
\begin{equation}
\label{new-s}
J\phi\rightarrow J\phi +J^1s_1\phi+J^2s_2\phi+J^Gs\phi,
\end{equation}
where the superscript $G$ means ``related to Grassmann numbers''. Now $J$ is replaced with the set $J^e=\{J,J^1,J^2,J^G\}$. This replace results in $W(J^e)$ depending on all the components of this extended $J^e$. Taking variational derivatives with respect to the extra components of $J^e$ generates the insertions $s_i\phi, s\phi$.

Now, the fact that the action of the system is invariant with respect to the transformation (\ref{transf}) implies that $W(J^e)$ is invariant with respect to the following transformation of the sources:
\begin{eqnarray}
\label{transf-s}
J&\rightarrow &J,  \nonumber\\
J^1&\rightarrow &J^1+J\epsilon^1,  \nonumber \\
J^2&\rightarrow &J^2+J\epsilon^2,  \nonumber \\
J^G&\rightarrow &J^G+J^i\epsilon^j\epsilon_{ji}.
\end{eqnarray}

Next, to reveal the content of the transformation (\ref{transf-s}), I introduce a source depending on two Grassmann variables $k_{1,2}$:
\begin{equation}
\label{grassmann-s}
\tilde{J}\equiv J+J^1k_1+J^2k_2+J^Gk_1k_2.
\end{equation}
Notice that $k_i$ have the mass dimension $+1$. They will become two extra momentum components.

With the new notations, the symmetry transformation (\ref{transf-s}), takes a simple form:
\begin{equation}
\label{grassmann-t}
\tilde{J}\rightarrow \tilde{J}+\tilde{J}\times(\epsilon^ik_i).
\end{equation}

Now, to further simplify the transformation (\ref{grassmann-t}), I introduce the Grassmann Fourier transform of $\tilde{J}$, depending on two Grassmann coordinates $\theta^{1,2}$:
\begin{equation}
\label{fourier}
J(\theta^1,\theta^2)\equiv\int dk_1dk_2\,\tilde{J}e^{\theta^ik_i}.
\end{equation}

With this notation, the symmetry transformation (\ref{grassmann-t}) implies the transformation
\begin{equation}
\label{final-t}
J(\theta^1,\theta^2)\rightarrow J(\theta^1+\epsilon^1,\theta^2+\epsilon^2).
\end{equation}

I see that $\theta^i$ can be interpreted as two extra (Grassmann) coordinates, and the theory is translation invariant with respect to shifts along these new coordinates. From now on, $J$ means the source depending on $\theta^i$, as defined in (\ref{fourier}).

Extending the number of space coordinates (even if the new coordinates are Grassmann numbers) comes not for free: we are to modify correspondingly the quantum Legendre transform. This is because now only the component of $\tilde{J}(k_1,k_2)$ at zero $k_i$ is the source for the fields. Correspondingly, only this component is involved in the quantum Legendre transform. I conclude that the inaction mapping $I_\mu$ breaks the translation invariance along the extra Grassmann coordinates. This constitutes the major complication: it is not easy to satisfy the inaction equation and the translation invariance along the Grassmann coordinates simultaneously. 

Now I have to define the symmetry generators $s_i$ used in (\ref{super-t-for-r}). To this end, I define $J_\epsilon(\theta^i)\equiv J(\theta^i+\epsilon^i)$. Then, the action of $s_i$ on any functional $W$ depending on $J$ is defined in the following way:
\begin{equation}
\label{def}
[s_iW](J)\equiv [\frac{\partial}{\partial\epsilon^i}W(J_\epsilon)]_{\epsilon^i=0}.
\end{equation} 

The last subject for this Section is how to solve Eq. (\ref{super-t-for-r}). Let us assume that any $R_\mu\in\mathcal{R}$ satisfying 
(\ref{super-t-for-r}) can be expanded in powers of a seed $S_\mu\in\mathcal{S}$:
\begin{equation}
\label{r-expansion}
R_\mu=S_\mu+\sum_{n=2}^\infty R_n(S_\mu),
\end{equation}
where $R_n(\lambda S_\mu)=\lambda^n R_n(S_\mu)$, and $R_n(S_\mu)\in\mathcal{R}$.

With this assumption, I deduce that
\begin{eqnarray}
\label{problem}
s_iR_2(S_\mu) & = & -s_iW_2(S_\mu)\nonumber\\
s_iR_n(S_\mu) & = & -s_i\Big[\sum_{k=2}^nW_k\Big((S_\mu+\dots+R_{n-1}(S_\mu)\Big)\Big]_n,
\end{eqnarray}
where the subscript $n$ in the right hand side means that only the terms of order $n$ in $S_\mu$ are retained inside the bracket.

The $W_k$ in the right hand sides of (\ref{problem}) do not belong to $\mathcal{R}$. Hopefully, it is possible to put them to $\mathcal{R}$ by adding vectors belonging to $\mathcal{I}$. This problem is still unsolved.

\section{\label{sec:out}Conclusions and outlook}

The inaction approach sketched above is self contained. It allows one to parameterize the Green functions of a theory in terms of a finite number of parameters---the coordinates in the finite dimensional linear seed space.

As demonstrated in \cite{gbp}, renormalization group equations can be formulated within the inaction approach for the parameters of the theory. The renormalization group equations of the inaction approach are not always equivalent to the standard renormalization group equations.

Now I want to point out unsolved problems. First of all, an exhaustive list of the super translation invariant theories satisfying the inaction equation should be given. Second, the equations (\ref{problem}) should be solved. The second problem is a technical one. Solving the first problem may give generalizations to gauge theories.

Let me give more comments on the problem of finding all the theories. Withing the inaction approach, to point out a theory means to point out a free theory $W_F\in\mathcal{P}$, and describe constructively the seed space corresponding to this $W_F$. 

There are known examples of such theories corresponding to the standard gauge theories. They possess extra properties: Lorentz invariance, and conservation of the ghost number.

Are there any other super translation invariant, local, and renormalizable theories? One can address this question within the inaction approach.

Optimistically, if such theories do exist, there may be the missing satisfactory and phenomenologically successful theory among them.

\end{document}